\documentclass[aps,superscriptaddress,showpacs,twocolumn]{revtex4}
\usepackage{graphicx}
\usepackage{color}
\usepackage{amssymb}
\usepackage{amsmath}

\newcommand{\bk}{{\bf k}}
\newcommand{\bq}{{\bf q}}
\newcommand{\bQ}{{\bf Q}}

\newcommand{\br}{{\bf r}}

\renewcommand{\Im}{{\mathop{\rm{Im}}\nolimits\,}}
\renewcommand{\Re}{{\mathop{\rm{Re}}\nolimits\,}}

\newcommand{\Tr}{{\mathop{\rm{Tr}}\nolimits\,}}
\newcommand{\EF}{E_{\mathrm{F}}}
\newcommand{\kB}{k_{\mathrm{B}}}
\newcommand{\Green}{{\mathcal G}}
\newcommand{\dSC}{{\mathrm{dSC}}}
\newcommand{\dPG}{{\mathrm{dPG}}}
\newcommand{\dDW}{{\mathrm{dDW}}}
\newcommand{\Ret}{{\mathrm{R}}}
\newcommand{\Tau}{T_\tau}

\begin{document}

\title{Linear response theory around a localized impurity in the
   pseudogap regime of an anisotropic superconductor: precursor pairing
   \emph{vs} the $d$-density-wave scenario}

\author{N. Andrenacci}
\email{Natascia.Andrenacci@unine.ch}
\affiliation{Institut de Physique, Universit\'e de Neuch\^atel,
   CH-2000 Neuch\^atel, Switzerland} 
\author{G. G. N. Angilella}
\email{Giuseppe.Angilella@ct.infn.it}
\affiliation{Dipartimento di Fisica e Astronomia, Universit\`a di
   Catania,\\ and Istituto Nazionale per la Fisica della Materia,
   UdR di Catania,\\ Via S. Sofia, 64, I-95123 Catania, Italy}
\author{H. Beck}
\affiliation{Institut de Physique, Universit\'e de Neuch\^atel,
   CH-2000 Neuch\^atel, Switzerland} 
\author{R. Pucci}
\affiliation{Dipartimento di Fisica e Astronomia, Universit\`a di
   Catania,\\ and Istituto Nazionale per la Fisica della Materia,
   UdR di Catania,\\ Via S. Sofia, 64, I-95123 Catania, Italy}

\date{\today}

\begin{abstract}
\medskip
We derive the polarizability of an electron system in (i) the
   superconducting phase, with $d$-wave symmetry, (ii) the pseudogap
   regime, within the precursor pairing scenario, and (iii) the
   $d$-density-wave (dDW) state, characterized by a $d$-wave hidden order
   parameter, but no pairing.
Such a calculation is motivated by the recent proposals that
   imaging the effects of an isolated impurity may distinguish between
   precursor pairing and dDW order in the pseudogap regime of the
   high-$T_c$ superconductors.
In all three cases, the wave-vector dependence of the polarizability
   is characterized by an azymuthal modulation, consistent with the
   $d$-wave symmetry of the underlying state.
However, only the dDW result shows the fingerprints of nesting, with
   nesting wave-vector $\bQ=(\pi,\pi)$, albeit imperfect, due to a
   nonzero value of the hopping ratio $t^\prime /t$ in the band
   dispersion relation.
As a consequence of nesting, the presence of hole pockets is also
   exhibited by the $(\bq,\omega)$ dependence of the retarded
   polarizability.
\\
\end{abstract} 

\maketitle

\section{Introduction}

Imaging of the electronic properties around an isolated
   nonmagnetic impurity such as Zn in the high-$T_c$ superconductors
   (HTS) has provided direct evidence of the unconventional nature of
   the superconducting state in the cuprates, and in particular of the
   $d$-wave symmetry of its order parameter below the critical
   temperature $T_c$ \cite{Pan:99,Hudson:99,Hudson:01,Pan:01}.
In the underdoped regime of the HTS, various models have
   been proposed to describe the pseudogap state above $T_c$.

Several experimental results provide substantial evidence of a pseudogap
   opening at the Fermi level in underdoped cuprates for $T_c < T <
   T^\ast$, even though no unique definition of the characteristic
   temperature $T^\ast$ is possible, as it generally depends
   on the actual experimental technique employed
   (see Refs.~\cite{Randeria:97-2,Timusk:99} for a review, and
   refs. therein). 
Also, the doping dependence of $T^\ast$ is still a matter of controversy
   \cite{Tallon:01}.
Owing to its $d$-wave symmetry, the pseudogap has been naturally
   interpreted in terms of precursor superconducting pairing.
In particular, the pseudogap has been associated to phase fluctuations
   of the order parameter above $T_c$ \cite{Emery:95a} (see
   Ref.~\cite{Loktev:01} for a review).
Within this precursor pairing scenario, the phase diagram of the HTS
   can be described as a crossover from Bose-Einstein
   condensation (in the underdoped regime) to BCS superconductivity
   (in the overdoped regime)
   \cite{Loktev:01,Randeria:95,Andrenacci:99,Strinati:00}. 

Recently, it has been proposed that many properties of the pseudogap
   regime may be explained within the framework of the so-called
   $d$-wave-density scenario (dDW)
   \cite{Chakravarty:01,Chakravarty:01a,Chakravarty:00}.
This is based on the idea that the pseudogap regime be characterized
   by a fully developed order parameter, at variance with the
   precursor pairing scenario, where a fluctuating order parameter is
   postulated.
The dDW state is an ordered state of unconventional kind, and is
   usually associated 
   with staggered orbital currents in the CuO$_2$ square lattice of
   the HTS \cite{Affleck:88,Marston:89,Kotliar:88,Schulz:89}.
Much attention has been recently devoted to show the consistency of
   the dDW scenario with several experimental properties of the HTS
   \cite{Chakravarty:01a}.
These include transport properties, such as the electrical and thermal
   conductivities \cite{Yang:02,Sharapov:03} and the Hall effect
   \cite{Chakravarty:02,Balakirev:03}, thermodynamic properties
   \cite{Kee:02,Wu:02}, time symmetry breaking \cite{Kaminski:02}, and
   angular resolved photoemission spectroscopy (ARPES)
   \cite{Chakravarty:03a}.
The possible occurrence of a dDW state in microscopic models of
   correlated electrons has been checked in ladder networks
   \cite{Marston:02}.

It has been recently proposed that
   direct imaging of the local density of states (LDOS) around an
   isolated impurity by means of scanning tunneling microscopy (STM)
   could help understanding the nature of the `normal' state in the
   pseudogap regime \cite{Zhu:01a,Wang:02,Morr:02,MoellerAndersen:03}.
The idea that an anisotropic superconducting gap should give rise to
   directly observable spatial features in the tunneling conductance
   near an impurity was suggested by Byers \emph{et al.}
   \cite{Byers:93}, whereas earlier studies \cite{Choi:90} had
   considered perturbations of the order parameter to occur within a
   distance of the order of the coherence length $\xi$ around an
   impurity.
Later, it was shown that an isolated impurity in a $d$-wave
   superconductor produces virtual bound states close to the Fermi
   level, in the nearly unitary limit \cite{Balatsky:95}.
Such a quasi-bound state should appear as a pronounced peak near the
   Fermi level in the LDOS at the impurity site \cite{Salkola:96}, as
   is indeed observed in Bi-2212 \cite{Pan:99} and YBCO
   \cite{Iavarone:02}.

In the normal state, the frequency-dependent LDOS at the nearest and
   next-nearest neighbor sites, with respect to the impurity site,
   should contain fingerprints of whether the pseudogap regime is
   characterized by precursor pairing \cite{Kruis:01} or dDW order
   \cite{Wang:02,Morr:02}. 
This is due to the fact that while pairing above $T_c$ without phase
   coherence is a precursor of Cooper pairing, and therefore of
   spontaneous breaking of U(1) gauge invariance, the dDW state can be
   thought as being characterized by the spontaneous breaking of
   particle-hole symmetry, in the same way as a charge density wave
   breaks pseudospin SU(2) symmetry \cite{Shen:97}.
The LDOS around a nonmagnetic impurity in both the dSC, the dDW and
   the competing dSC+dDW phases in the underdoped regime has been
   actually calculated \emph{e.g.} by Zhu \emph{et al.}
   \cite{Zhu:01a}.

In this context, a complementary information is that provided by the
   polarizability $F^\Ret (\bq,\omega)$ of the system, which gives a
   measure of the linear response of the charge density to an impurity
   potential.
In the case of $d$-wave superconductors, it has been demonstrated that
   the anisotropic dependence of the superconducting order parameter
   on the wave-vector $\bq$ gives rise to a clover-like azymuthal
   modulation of $F^\Ret (\bq,\omega)$ along the Fermi line for a 2D
   system \cite{Angilella:02f}.

These patterns in the $\bq$ dependence of $F^\Ret (\bq,\omega)$ are
   here confirmed also for a more realistic band for the cuprates.
In addition to that, the dDW result also shows fingerprints of the
   $\bQ=(\pi,\pi)$ nesting properties of such a state.

The paper is organized as follows.
In Sec.~\ref{sec:dSCdPG}, we review the expression of the
   polarizability for the $d$-wave superconducting state (dSC) and
   derive that of the $d$-wave pseudogap regime, within the precursor
   pairing scenario (dPG).
In Sec.~\ref{sec:dDW}, we derive the polarizability for the dDW state.
By allowing nonzero values of the hopping ratio $t^\prime /t$ in the
   dispersion relation \cite{Pavarini:01,Angilella:01,Angilella:03g},
   we will explicitly consider the case in which perfect nesting is
   destroyed.
Such a case is relevant for the study of the dDW state, given its
   particle-hole character.
In Sec.~\ref{sec:dSCdDW}, we consider the competition of
   dDW order with an subdominant dSC state in the underdoping regime.
In Sec.~\ref{sec:numerical}, we present our numerical results for the
   polarizability in the dSC, dPG, and dDW states, both in the static
   limit and as a function of frequency.
We eventually summarize and make some concluding remarks in
   Sec.~\ref{sec:conclusions}.

\section{Linear response function in the \lowercase{d}SC and
   \lowercase{d}PG states} 
\label{sec:dSCdPG}

Within linear response theory, the displaced charge density
   $\delta\rho(\br)$ by a scattering potential $V(\br)$ in the Born
   approximation is given by
\begin{equation}
\delta\rho(\br) = \int  V(\br^\prime ) F^\Ret (\br-\br^\prime
   ,\EF ) d\br^\prime ,
\label{eq:lr}
\end{equation}
which implicitly defines the linear response function $F^\Ret
   (\br,\EF)$ at the Fermi energy $\EF$.
Here and in the following we set the elementary charge $e=1$.
Its relevance in establishing the electronic structure of isolated
   impurities in normal metals and alloys has been earlier emphasized
   by Stoddart \emph{et al.} \cite{Stoddart:69,Jones:73-2}.
In momentum space, Eq.~(\ref{eq:lr}) readily translates into $\delta
   \rho(\bq) = V(\bq) F^\Ret (\bq,\EF)$, showing that, for a highly
   localized scattering potential in real space [$V(\br)=V_0
   \delta(\br)$, say], the Fourier transform $\delta\rho(\bq)$ of the
   displaced charge is simply proportional to $F^\Ret (\bq,\EF)$.

In the presence of superconducting pairing, the generalization of the
   linear response function is given by the density-density
   correlation function (polarizability) \cite{Prange:63}: 
\begin{eqnarray}
F(\bq,i\omega_\nu ) &=& \Tr \frac{1}{\beta} \sum_{\omega_n} \frac{1}{N} \sum_\bk
   \tau_3 \Green (\bk,i\omega_n ) \nonumber\\
&&\times \, \tau_3 \Green (\bk-\bq,i\omega_n
   -i\omega_\nu ) 
\label{eq:Prange}
\end{eqnarray}
where $\Green (\bk,i\omega_n )$ is the matrix Green's function in
   Nambu notation, $\beta=T^{-1}$ is the inverse temperature,
   $\omega_\nu = 2\nu\pi T$ is a bosonic Matsubara frequency, $\tau_i$ are the
   Pauli matrices in spinor space, the summations are performed over the $N$
   wave-vectors $\bk$ of
   the first Brillouin zone (1BZ) and all fermionic Matsubara
   frequencies $\omega_n = (2n+1)\pi T$, and the trace is over the
   spin indices.
Here and below we shall use units such that $\hbar = \kB
   = 1$ and lattice spacing $a=1$.
The retarded polarizability is defined as usual in terms of the
   analytic continuation as $F^\Ret (\bq,\omega) =
   F(\bq,i\omega_\nu \mapsto \omega+i0^+ )$.
In the normal state, Equation~(\ref{eq:Prange}) correctly reduces to
   the Lindhard function for the polarizability of a free electron gas
   \cite{Mahan:90}. 

In the following, by specifying the functional form of $\Green$ in
   the case of pairing with and without phase coherence, we will in turn
   derive the explicit expression for $F$ in the superconducting phase
   with a $d$-wave order parameter,
   and in the pseudogap regime, characterized by fluctuating
   $d$-wave order (precursor pairing scenario).

\subsection{Superconducting phase}

We assume the following BCS-like Hamiltonian:
\begin{equation}
H_\dSC = \sum_{\bk s} \xi_\bk c^\dag_{\bk s}
   c_{\bk s} + \sum_{\bk\bk^\prime} V_{\bk\bk^\prime}
   c^\dag_{\bk\uparrow} c^\dag_{-\bk\downarrow} c_{-\bk^\prime
   \downarrow} c_{\bk^\prime \uparrow} ,
\label{eq:HdSC}
\end{equation}
where $c^\dag_{\bk s}$ ($c_{\bk s}$) is a creation (annihilation)
   operator for an electron in the state with wave-vector $\bk$ and
   spin projection $s\in\{\uparrow,\downarrow\}$ along a specified
   direction, and $\xi_\bk = \epsilon_\bk -\mu$, with $\epsilon_\bk$
   the single-particle dispersion relation: 
\begin{equation}
\epsilon_\bk = -2t(\cos k_x + \cos k_y ) + 4t^\prime \cos k_x \cos k_y
   ,
\label{eq:disp}
\end{equation}
where $t=0.3$~eV, $t^\prime /t =0.3$ are tight binding
   hopping parameters 
   appropriate for the cuprate superconductors, and $\mu$ the chemical
   potential.
In Eq.~(\ref{eq:HdSC}), $V_{\bk\bk^\prime}$ is a model
   potential, which we assume to be separable and attractive in the
   $d_{x^2 -y^2}$-wave channel: $V_{\bk\bk^\prime} = -\lambda g_\bk
   g_{\bk^\prime}$, with $g_\bk = \frac{1}{2} (\cos k_x - \cos k_y )$
   and $\lambda>0$.
Under these assumptions, the Hamiltonian, Eq.~(\ref{eq:HdSC}), is
   characterized by a nonzero superconducting order parameter $\langle
   c_{\bk\uparrow} c_{-\bk\downarrow} \rangle$, leading to a nonzero
   $d$-wave mean-field gap $\Delta_\bk = \Delta_\circ g_\bk$ below the
   critical temperature $T_c$.

Making use of the explicit expression for the matrix Green's function
   $\Green_\dSC$ in the superconducting state \cite{AGD}:
\begin{equation}
\Green_\dSC (\bk,i\omega_n ) = \frac{i\omega_n \tau_0
   +\xi_\bk
   \tau_3 +\Delta_\bk \tau_1}{(i\omega_n )^2 - E_\bk^2 },
\label{eq:GdSC}
\end{equation}
with $E_\bk = (\xi_\bk^2 + \Delta_\bk^2 )^{1/2}$ the
   upper branch of the superconducting spectrum and $\tau_0$ the
   identity matrix in spin space, and performing the
   trace over spin indices
   and the summation over the internal frequency \cite{Mahan:90} in
   Eq.~(\ref{eq:Prange}), 
   we obtain the linear response function for a $d$-wave
   superconducting state \cite{Prange:63}:
\begin{widetext}
\begin{eqnarray}
F_\dSC (\bq,i\omega_\nu ) &=&
\frac{1}{N} \sum_\bk \left[
(u_\bk u_{\bk-\bq} - v_\bk v_{\bk-\bq} )^2 \left(
\frac{f(E_\bk ) - f(E_{\bk-\bq} )}{E_\bk -E_{\bk-\bq} -i\omega_\nu}
+ \mathrm{H.c.} \right) \right.\nonumber \\
&&\left. +
(u_{\bk-\bq} v_\bk + u_\bk v_{\bk-\bq} )^2 \left(
\frac{f(E_\bk ) + f(E_{\bk-\bq} )-1}{E_\bk +E_{\bk-\bq} -i\omega_\nu}
+ \mathrm{H.c.} \right) \right],
\label{eq:FdSC}
\end{eqnarray}
\end{widetext}
where $u_\bk^2 =
   \frac{1}{2} [1+ \xi_\bk /E_\bk ]$, $v_\bk^2 = 1-u_\bk^2$
   are the usual coherence factors of BCS theory, and
   $f(\epsilon)=(1+e^{\beta\epsilon} )^{-1}$ is the 
   Fermi function at temperature $T$.
In the limit of zero external frequency and $T\to0$,
   Eq.~(\ref{eq:FdSC}) reduces to the static polarizability studied in
   Ref.~\cite{Angilella:02f} for a $d$-wave superconductor.

\subsection{Pseudogap regime, within the precursor pairing scenario}

In the pseudogap regime, for $T_c < T <T^\ast$, within the precursor
   pairing scenario \cite{Loktev:01}, one assumes the existence of
   Cooper pairs characterized by a `binding energy' $\Delta_\bk$
   having the same symmetry of the true 
   superconducting gap below $T_c$, but no phase coherence.
In other words, no true off-diagonal long range order develops, and
   one rather speaks of a `fluctuating' order \cite{Emery:95a}.
This means that the quasiparticle spectrum $E_\bk = (\xi_\bk^2 +
   \Delta_\bk^2 )^{1/2}$ is still characterized by a pseudogap
   $\Delta_\bk = \Delta_\circ g_\bk$ opening at the Fermi energy with
   $d$-wave symmetry, but now without phase coherence.
Therefore, the diagonal elements of the matrix Green's function
   $\Green_\dPG$ coincide with those of its superconducting
   counterpart, Eq.~(\ref{eq:GdSC}), while the off-diagonal, anomalous
   elements are null:
\begin{equation}
\Green_\dPG (\bk,i\omega_n ) = \frac{i\omega_n \tau_0
   +\xi_\bk \tau_3 }{(i\omega_n )^2 - E_\bk^2 }.
\label{eq:GdPG}
\end{equation}
The effects due to a finite lifetime of the precursor Cooper
   pairs can be mimicked by adding a finite imaginary energy linewidth
   $i\Gamma$ to the dispersion relation entering Eq.~(\ref{eq:GdPG}),
   or by substituting the spectral functions associated with the
   quasiparticle states with `broadened' ones, as discussed in
   Appendix~\ref{app:lifetime}.
The relation between the two approaches and with analytical
   continuation has been discussed in Appendix~A of
   Ref.~\cite{Andrenacci:03}.

Within this precursor pairing scenario, Equation~(\ref{eq:Prange}) in the
   pseudogap regime then reads:
\begin{widetext}
\begin{eqnarray}
F_\dPG (\bq,i\omega_\nu ) &=&
\frac{1}{N} \sum_\bk \left[
(u_\bk^2 u_{\bk-\bq}^2 + v_\bk^2 v_{\bk-\bq}^2 ) \left(
\frac{f(E_\bk ) - f(E_{\bk-\bq} )}{E_\bk -E_{\bk-\bq} -i\omega_\nu}
+ \mathrm{H.c.} \right) \right.\nonumber \\
&&\left. +
(u_{\bk-\bq}^2 v_\bk^2 + u_\bk^2 v_{\bk-\bq}^2 ) \left(
\frac{f(E_\bk ) + f(E_{\bk-\bq} )-1}{E_\bk +E_{\bk-\bq} -i\omega_\nu}
+ \mathrm{H.c.} \right) \right],
\label{eq:FdPG}
\end{eqnarray}
\end{widetext}
where we are implicitly assuming $T_c < T < T^\ast$.

\section{Linear response function in the \lowercase{d}DW state} 
\label{sec:dDW}

The mean-field Hamiltonian for the $d$-density-wave state is
   \cite{Chakravarty:01}:
\begin{equation}
H_\dDW = \sum_{\bk s} [\xi_\bk c^\dag_{\bk s}
   c_{\bk s} + i D_\bk c^\dag_{\bk s} c_{\bk+\bQ s} ],
\label{eq:HdDW}
\end{equation}
where the summation is here restricted to all wave-vectors $\bk$ belonging
   to the first Brillouin zone, $\bQ = (\pi,\pi )$ is the dDW ordering
   wave-vector, and $D_\bk = D_\circ g_\bk$ is the dDW order
   parameter.
As anticipated above, the dDW state is characterized by a broken
   symmetry and a well-developed order parameter, at variance with the
   precursor pairing scenario of the pseudogap regime.
Such a state is associated to staggered orbital currents circulating
   with alternating sense in the neighboring plaquettes of the
   underlying square lattice.
As a result, the unit cell in real space is doubled, and the Brillouin
   zone is correspondingly halved.
At variance with other `density waves', the dDW order is characterized
   not by charge or spin modulations, but rather by current
   modulations.

The nonzero, singlet order parameter $\Phi_\bQ$ breaks pseudospin
   invariance in the particle-hole space:
\begin{equation}
\langle c^\dag_{\bk+\bQ s} c_{\bk s^\prime} \rangle = i \Phi_\bQ g_\bk
   \delta_{ss^\prime} .
\label{eq:dDWop}
\end{equation}
Whereas it possesses $d$-wave symmetry, as expected, its imaginary
   value leads to the breaking of a relatively large number of
   symmetries, such as time reversal, parity, translation by a lattice
   spacing, and rotation by $\pi/2$, although the product of any two
   of these is preserved (see Ref.~\cite{Sharapov:03} for a detailed
   analysis).

Introducing the spinor $\Psi_{\bk s}^\dag = (c^\dag_{\bk s} \,\,
   c^\dag_{\bk+\bQ s} )$, the dDW Hamiltonian, Eq.~(\ref{eq:HdDW}), can be
   conveniently rewritten as \cite{Yang:02,Sharapov:03}:
\begin{equation}
H_\dDW = {\sum_{\bk s}}^\prime \Psi_{\bk s}^\dag [(\epsilon_\bk^+
   -\mu ) \tau_0 + \epsilon_\bk^- \tau_3 + D_\bk \tau_1 ] \Psi_{\bk s} ,
\end{equation}
where $\epsilon_\bk^\pm = \frac{1}{2} (\epsilon_\bk \pm
   \epsilon_{\bk+\bQ} )$, and the prime restricts the summation over
   wave-vectors $\bk$ belonging to the reduced (`magnetic') Brillouin
   zone only.
Notice that $\epsilon^\pm_{\bk+\bQ} = \mp\epsilon_\bk^\pm$.
Correspondingly, the matrix Green's function at the imaginary time
   $\tau$ can be defined as $\Green_\dDW (\bk,\tau) = -\langle \Tau
   \Psi_{\bk s} (\tau) \Psi^\dag_{\bk s} (0) \rangle$, whose inverse reads
   \cite{Morr:02,Sharapov:03}: 
\begin{equation}
\Green_\dDW^{-1} (\bk,i\omega_n ) =
\begin{pmatrix}
i\omega_n -\xi_\bk & iD_\bk \\
-iD_\bk & i\omega_n -\xi_{\bk+\bQ} 
\end{pmatrix}.
\label{eq:iGdDW}
\end{equation}
In the case of perfect nesting ($t^\prime = 0$) for the dispersion
   relation, Eq.~(\ref{eq:disp}), Sharapov \emph{et al.}
   \cite{Sharapov:03} explicitly find
\begin{equation}
\Green_\dDW (\bk,i\omega_n ) = \frac{(i\omega_n + \mu)\tau_0 +
   \epsilon_\bk \tau_3 -D_\bk \tau_2}{(i\omega_n + \mu)^2 -
   \epsilon_\bk^2 - D_\bk^2} ,
\label{eq:GdDWSharapov}
\end{equation}
to be compared and contrasted with Eq.~(\ref{eq:GdSC}) for the
   superconducting phase.
Notice, in particular, the different way the chemical potential $\mu$
   enters the two expressions.

In the general case ($t^\prime \neq 0$), perfect nesting is lost, and
   we have to refer to the general form of $\Green_\dDW^{-1}$,
   Eq.~(\ref{eq:GdDW}).
One finds:
\begin{eqnarray}
\Green_\dDW (\bk,i\omega_n ) &=&
\frac{1}{(i\omega_n -E_\bk^+ )(i\omega_n -E_\bk^- )}
   \nonumber \\
&&\times\,\begin{pmatrix}
i\omega_n -\xi_{\bk+\bQ} & -iD_\bk \\
iD_\bk & i\omega_n -\xi_\bk
\end{pmatrix},
\label{eq:GdDW}
\end{eqnarray}
where $E_\bk^\pm = -\mu + \epsilon_\bk^+ \pm \sqrt{(\epsilon_\bk^- )^2
   +D_\bk^2}$ are the two branches of the quasiparticle spectrum
   obtained by diagonalizing Eq.~(\ref{eq:HdDW}) \cite{Yang:02}.
Notice that $E_{\bk+\bQ}^\pm = E^\pm_\bk$.

In the limit $t^\prime =0$, Eq.~(\ref{eq:GdDW}) correctly reduces to
   Eq.~(\ref{eq:GdDWSharapov}), even though it is not straightforward
   to express Eq.~(\ref{eq:GdDW}) in the same compact matrix notation.
In the limit of perfect nesting ($t^\prime =0$), the dispersion
   relation, Eq.~(\ref{eq:disp}), is antisymmetric with respect to
   particle-hole conjugation, $\epsilon_{\bk+\bQ} = -\epsilon_\bk$.
As a result, $E_\bk^\pm = -\mu \pm (\epsilon_\bk^2 + D_\bk^2 )^{1/2}$,
   which is to be contrasted with the quasiparticle spectrum of the
   superconducting state or the pseudogap state within the precursor
   pairing scenario, $\pm E_\bk = \pm [(\epsilon_\bk -\mu)^2 +
   \Delta_\bk^2 ]^{1/2}$.
The difference comes again from the fact that the Bogoliubov
   excitations in the dSC and the dPG states are Cooper pairs, while
   the dDW ordered state is characterized by particle-hole mixture
   \cite{Kee:02,MoellerAndersen:03}. 

The polarizability in the dDW state is derived in
   Appendix~\ref{app:poldDW}.
We just quote here the final result, which can be cast in compact
   matrix notation as:
\begin{eqnarray}
F_\dDW (\bq,i\omega_\nu ) &=&
\Tr \frac{1}{\beta} \sum_{\omega_n} \frac{1}{N} {\sum_\bk}^\prime
\kappa \Green_\dDW (\bk,i\omega_n )
\nonumber\\
&&\times \, \kappa \Green_\dDW (\bk-\bq,i\omega_n -i\omega_\nu ),
\label{eq:PrangedDW}
\end{eqnarray}
where now $\kappa = \tau_0 + \tau_1$, and $\Green_\dDW$ is
   the matrix Green's function for the dDW state, Eq.~(\ref{eq:GdDW}).
Performing the frequency summation \cite{Mahan:90}, one eventually
   finds:
\begin{equation}
F_\dDW (\bq,i\omega_\nu ) = \frac{1}{N} {\sum_\bk}^\prime
\sum_{i,j=\pm} \frac{f(E_\bk^i ) - f(E_{\bk-\bq}^j )}{E_\bk^i -
   E_{\bk-\bq}^j -i\omega_\nu} .
\label{eq:FdDW}
\end{equation}

\section{Competition between \lowercase{d}\uppercase{SC} and
   \lowercase{d}\uppercase{DW} orders}
\label{sec:dSCdDW}

In the underdoped regime, it has been predicted on phenomenological
   grounds that the dDW order should compete with a subdominant dSC phase
   \cite{Chakravarty:01}.
This has been confirmed by model calculations at the mean-field level
   \cite{Zhu:01,Wu:02}, showing that indeed an existing broken
   symmetry of dDW kind at high temperature suppresses that critical
   temperature for the subdominant dSC ordered phase.
Recently, the competition between dDW and dSC orders has been shown to
   be in agreement with the unusual $T$-dependence of the restricted
   optical sum rule, as observed in the underdoped HTS
   \cite{Benfatto:03}.

In order to take into account for the competition between the dSC and dDW
   orders at finite temperature, one has to separately consider the
   electron states within the two inequivalent halves of the Brillouin
   zone.
Therefore, it is convenient to make use of the 4-components Nambu
   spinor $\Psi_\bk^\dag \equiv (\Psi_{\bk\uparrow}^\dag ~~
   \Psi_{-\bk\downarrow}^\top )$, or explicitly:
\begin{equation}
\Psi_\bk = \begin{pmatrix}
c_{\bk \uparrow} \\
c_{\bk+\bQ  \uparrow} \\
c^\dag_{-\bk \downarrow} \\
c^\dag_{-\bk-\bQ \downarrow}
\end{pmatrix} .
\end{equation}
At the mean-field level, the Hamiltonian for the competing dSC and
   dDW phases thus reads
\begin{equation}
H_{\dSC + \dDW} = {\sum_\bk}^\prime \Psi_\bk^\dag \hat{H}_\bk \Psi_\bk ,
\end{equation}
where $\hat{H}_\bk$ is the $4\times 4$ Hermitean matrix defined by
\begin{equation}
\hat{H}_\bk = \begin{pmatrix}
\xi_\bk & iD_\bk & \Delta_\bk & 0 \\
-iD_\bk & \xi_{\bk+\bQ} & 0 & -\Delta_\bk \\
\Delta_\bk^\ast & 0 & -\xi_{-\bk} & iD_{-\bk} \\
0 & -\Delta^\ast_\bk & -iD_{-\bk} & - \xi_{-\bk-\bQ} 
\end{pmatrix},
\label{eq:Hmatrix}
\end{equation}
with real eigenvalues $E_{\bk i}$ (here, $i=0,\ldots 3$) given by
\begin{eqnarray}
E^2_{\bk i} &=& \Delta_\bk^2 + D_\bk^2 - \xi_\bk \xi_{\bk + \bQ}
\nonumber\\
\!\!\!\!\!
&&+ (\xi_\bk + \xi_{\bk + \bQ} ) 
\left[
\epsilon_\bk^+ -\mu \pm \sqrt{(\epsilon_\bk^- )^2 +D_\bk^2}
\right],
\label{eq:Eki}
\end{eqnarray}
and orthonormal eigenvectors $|\bk i\rangle$ (for given $\bk$ in
   the reduced 1BZ).
It may be straightforwardly checked that $E_{\bk i}$ reduces to the dSC
   superconducting spectrum, $\pm E_\bk$, and to the dDW quasiparticle
   dispersion 
   relations, $E_\bk^\pm$, in the limits $D_\circ =0$ (pure dSC) and
   $\Delta_\circ = 0$ (pure dDW), respectively, when the halving of
   the 1BZ is removed.

In the particle-hole symmetric case ($t^\prime = 0$), making use of
   the nesting properties described in Sec.~\ref{sec:dDW},
   Eq.~(\ref{eq:Eki}) simplifies as:
\begin{equation}
E_{\bk i} = \pm \sqrt{\Delta_\bk^2 + \left( \mu \pm
   \sqrt{\epsilon_\bk^2 + D^2_\bk } \right)^2} .
\end{equation}
For $\mu=0$, the four branches of the spectrum degenerate into the Dirac cone
   \cite{Lee:97,Lee:97a}
\begin{equation}
E_{\bk,0\equiv 2,1\equiv 3} = \pm \sqrt{\epsilon_\bk^2 + \Delta_\bk^2
   + D^2_\bk} ,
\label{eq:cones}
\end{equation}
thus showing that the two gaps have the same role, \emph{i.e.} the
   system may be equivalently described as a dDW or a dSC
   superconductor, with a $d$-wave gap $\sqrt{\Delta_\bk^2 + D^2_\bk}$
   in either case.
Either a non-zero hopping ratio ($t^\prime /t \neq 0$) or a hole-doping away
   from half-filling ($\mu\neq 0$ when $t^\prime = 0$) destroys this
   particular symmetry, and one has to resort to the eigenvalues
   $E_{\bk i}$, Eq.~(\ref{eq:Eki}).

In order to obtain the Green's functions in the dSC+dDW case, it is
   useful to introduce the $4\times 4$ matrices ($i,j=0,\ldots 3$)
\begin{equation}
\Gamma_{ij} = \tau_i \otimes \tau_j ,
\end{equation}
whose algebra is given by
\begin{equation}
\Gamma_{ij} \Gamma_{lm} = i \varepsilon_{ijk} i \varepsilon_{lmn}
   \Gamma_{kn} ,
\end{equation}
where $\varepsilon_{ijk}$ is the totally antisymmetric Levi-Civita
   tensor, and
\begin{equation}
\Gamma_{33}^\pm = \frac{1}{2} (\Gamma_{30} \pm \Gamma_{33} ).
\end{equation}
Then the matrix Hamiltonian Eq.~(\ref{eq:Hmatrix}) takes the form
\begin{equation}
\hat{H}_\bk = \epsilon_\bk \Gamma_{33}^+ + \epsilon_{\bk+\bQ}
   \Gamma_{33}^- - \mu \Gamma_{30} - D_\bk \Gamma_{02} + \Delta_\bk
   \Gamma_{13} ,
\end{equation}
whence the inverse Green's function (now a $4\times 4$ matrix in Nambu
   space) straightforwardly follows as 
\begin{equation}
\Green^{-1}_{\dSC+\dDW} (\bk ,i\omega_n ) =
i\omega_n \Gamma_{00} - \hat{H}_\bk .
\end{equation}
As in the dDW case, Eq.~(\ref{eq:PrangedDW}) (see also
   Appendix~\ref{app:poldDW}), the linear response function in the
   dSC+dDW case can be given a compact matrix form as
\begin{eqnarray}
F_{\dSC+\dDW} (\bq,i\omega_\nu ) &=&
\Tr \frac{1}{\beta} \sum_{\omega_n} \frac{1}{N} {\sum_\bk}^\prime
\hat{\kappa} \Green_{\dSC+\dDW} (\bk,i\omega_n )
\nonumber\\
&&\!\!\!\!\!\!\!
\times \, \hat{\kappa} \Green_{\dSC+\dDW} (\bk-\bq,i\omega_n -i\omega_\nu ),
\label{eq:PrangedSCdDW}
\end{eqnarray}
where now the vertex matrix in the $4\times 4$ Nambu spinor space is
   given by
\begin{equation}
\hat{\kappa} = \tau_3 \otimes \kappa = \begin{pmatrix} \tau_0 + \tau_1 & 0
   \\
0 & -\tau_0 -\tau_1 \end{pmatrix} .
\end{equation}
Finally, it can be shown that Eq.~(\ref{eq:PrangedSCdDW}) also admits
   the following spectral decomposition, analogous to
   Eq.~(\ref{eq:FdDW}):
\begin{eqnarray}
F_{\dSC+\dDW} (\bq,i\omega_\nu ) &=& \nonumber \\
&&\!\!\!\!\!\!\!\!\!\!\!\!\!\!\!\!\!\!\!\!\!
\frac{1}{N} {\sum_\bk}^\prime
\sum_{i,j} \frac{f(E_{\bk i} ) - f(E_{\bk-\bq j}
   )}{E_{\bk i} - E_{\bk-\bq j} -i\omega_\nu} u_{ij}
   (\bk,\bq) , \nonumber \\
\label{eq:FdSCdDW}
\end{eqnarray}
where $u_{ij} (\bk,\bq) = \Tr (\hat{\kappa} P_{\bk i} \hat{\kappa}
   P_{\bk-\bq j} )$, and $P_{\bk i} =
   |\bk i\rangle\langle\bk i|$ is the 
   orthonormal projector operator on the $i$ eigenstate of the
   matrix Hamiltonian, Eq.~(\ref{eq:Hmatrix}).

\section{Numerical results and discussion}
\label{sec:numerical}

We have evaluated numerically the polarizability for the dPG and the
   dDW cases, Eqs.~(\ref{eq:FdPG}) and (\ref{eq:FdDW}), and
   for the mixed dSC+dDW case, Eq.~(\ref{eq:FdSCdDW}), as a function of
   the relevant variables. 
Our numerical results for the pure dSC case turn out to be
   very similar to 
   the dPG case (at least over the range of variables considered
   below), and will not be shown here.
In the dPG and in the pure dDW cases,
we adopt the following set of parameters, which are believed to be
   relevant for the cuprate superconductors: $t=0.3$~eV, $t^\prime /t
   = 0.3$, $\mu=-t$, corresponding to a hole-like Fermi line and a
   hole doping $\sim 14.3\%$, $\Delta_\circ = D_\circ = 0.06$~eV 
   in the dPG and in the dDW cases, respectively
   \cite{Chakravarty:03a}, and $T=100$~K.

\begin{figure}[t]
\begin{center}
\begin{minipage}[c]{0.49\columnwidth}
\includegraphics[height=\textwidth,angle=-90]{lrpg_pgplot_01.ps}
\end{minipage}
\begin{minipage}[c]{0.49\columnwidth}
\includegraphics[height=1.45\textwidth,angle=-90]{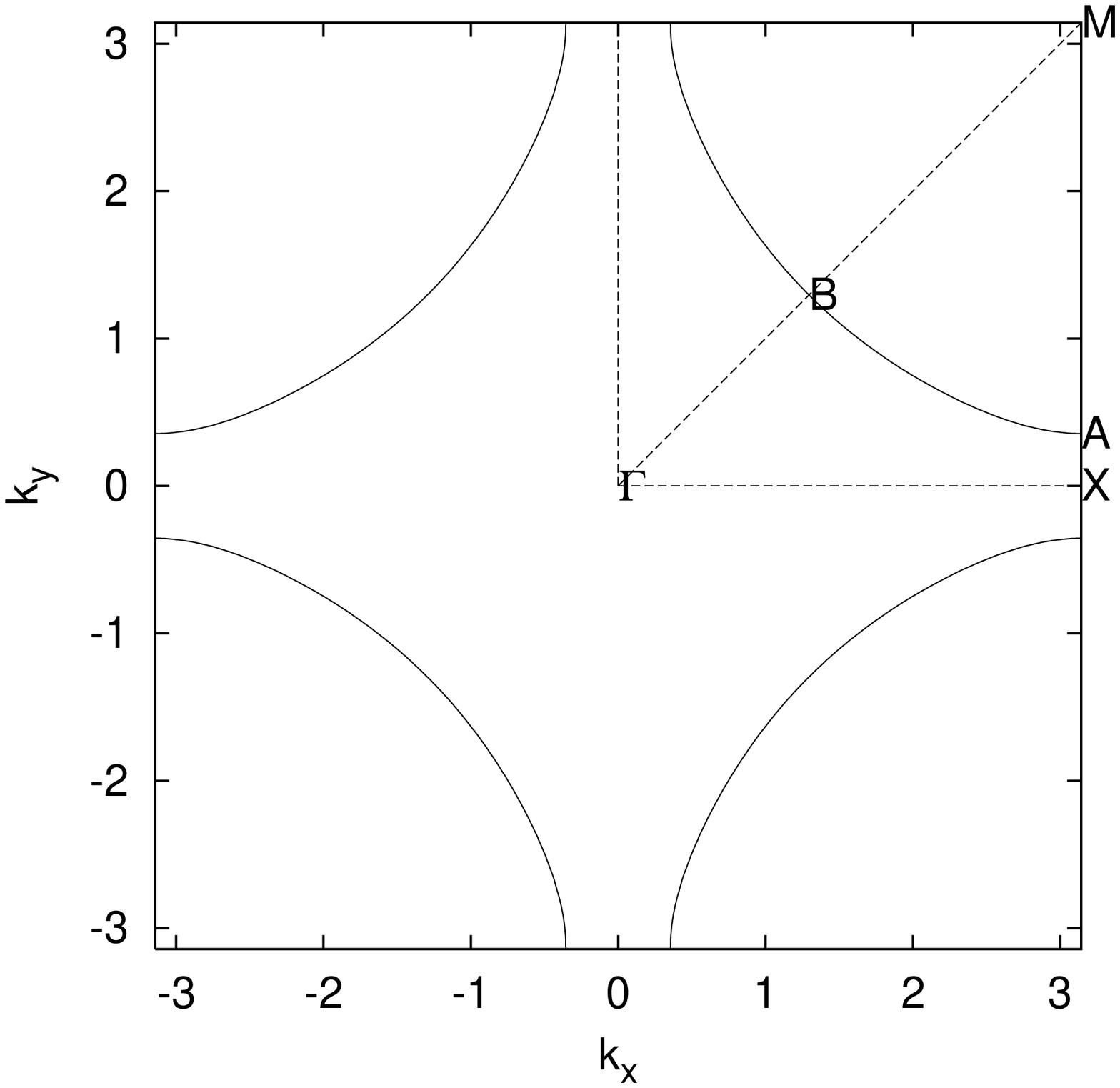}
\end{minipage}\\
\begin{minipage}[c]{0.49\columnwidth}
\begin{flushleft}
(a)
\end{flushleft}
\end{minipage}
\begin{minipage}[c]{0.49\columnwidth}
\begin{flushleft}
(b)
\end{flushleft}
\end{minipage}
\end{center}
\caption{Static polarizability for the dPG state in momentum
   space, $F_\dPG(\bq,0)$, Eq.~(\protect\ref{eq:FdPG}) [in eV$^{-1}$,
   panel (a)], for $\Delta_\circ = 0.06$~eV, $T=100$~K, and $\mu=-t =
   -0.3$~eV, corresponding to the hole-like 
   Fermi line in the normal state shown in panel (b).
Panel (b) also reports the special points $\Gamma=(0,0)$,
   $X=(\pi,0)$, $M=(\pi,\pi)$, along with the points $A$ and $B$ where
   the Fermi line $\xi_\bk = 0$ intersects the
   symmetry contour $\Gamma$--$X$--$M$--$\Gamma$.
}
\label{fig:qdPG}
\end{figure}

\begin{figure}[t]
\begin{center}
\begin{minipage}[c]{0.49\columnwidth}
\includegraphics[height=\textwidth,angle=-90]{lrddw_pgplot_01.ps}
\end{minipage}
\begin{minipage}[c]{0.49\columnwidth}
\includegraphics[height=1.45\textwidth,angle=-90]{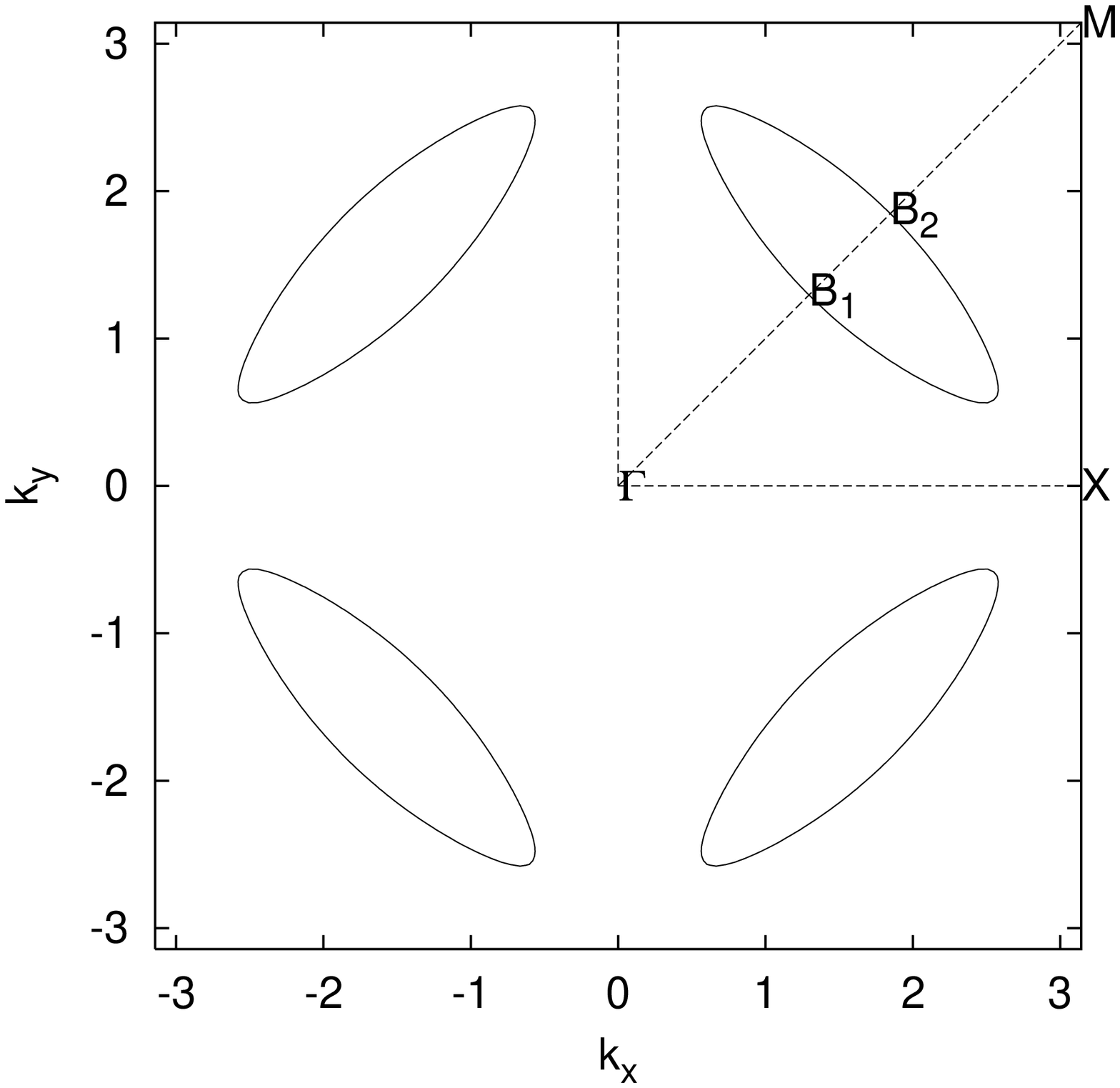}
\end{minipage}\\
\begin{minipage}[c]{0.49\columnwidth}
\begin{flushleft}
(a)
\end{flushleft}
\end{minipage}
\begin{minipage}[c]{0.49\columnwidth}
\begin{flushleft}
(b)
\end{flushleft}
\end{minipage}
\end{center}
\caption{Static polarizability for the pure dDW state in momentum
   space, $F_\dDW(\bq,0)$, Eq.~(\protect\ref{eq:FdDW}) [in eV$^{-1}$,
   panel (a)], for $D_\circ = 0.06$~eV, $T=100$~K, and
   $\mu=-t=-0.3$~eV, corresponding to the hole pockets 
   $E_\bk^- = 0$ shown in panel (b).
Panel (b) also reports the special points $\Gamma=(0,0)$,
   $X=(\pi,0)$, $M=(\pi,\pi)$, along with the points $B_1$ and $B_2$ where
   the line $E_\bk^- = 0$ intersects the
   symmetry contour $\Gamma$--$X$--$M$--$\Gamma$.
}
\label{fig:qdDW}
\end{figure}

\subsection{Zero external frequency}
\label{ssec:static}

In order to make contact with earlier work \cite{Angilella:02f}, we
   first consider the case of zero external (bosonic) frequency,
   $\omega_\nu = 0$, in the time-ordered polarizabilities,
   Eqs.~(\ref{eq:FdPG}) and (\ref{eq:FdDW}).

Our numerical results for the wave-vector dependence of $F(\bq,0)$
   over the 1BZ in the dPG and in the pure dDW cases are shown in
   Figs.~\ref{fig:qdPG} and \ref{fig:qdDW}, respectively.
As a result of the $d$-wave symmetry of both the pseudogap within the
   precursor pairing scenario, and of the dDW order parameter, $F(\bq,0)$
   is characterized by a four-lobed pattern or azymuthal modulation
   \cite{Angilella:02f}.
However, the dDW case is also characterized by the presence of `hole
   pockets', centered around $\bQ/2 = (\pi/2,\pi/2)$ and symmetry related
   points, due to the (albeit 
   imperfect) nesting properties of the dDW state, with nesting vector
   $\bQ=(\pi,\pi)$ (see Fig.~\ref{fig:qdDW}b).
Such a feature is reflected in the $\bq$ dependence of $F_\dDW
   (\bq,0)$, which is characterized by local maxima at the hole
   pockets, for the value $\mu=-t$ of the chemical potential
   considered here.
(Other values of the chemical potentials give rise to analogous
   features, which are absent in the dPG case.)

\begin{figure}[t]
\begin{center}
\begin{minipage}[c]{0.49\columnwidth}
\includegraphics[height=\textwidth,angle=-90]{lrmixedr_pgplot_01.ps}
\end{minipage}
\begin{minipage}[c]{0.49\columnwidth}
\includegraphics[height=1.45\textwidth,angle=-90]{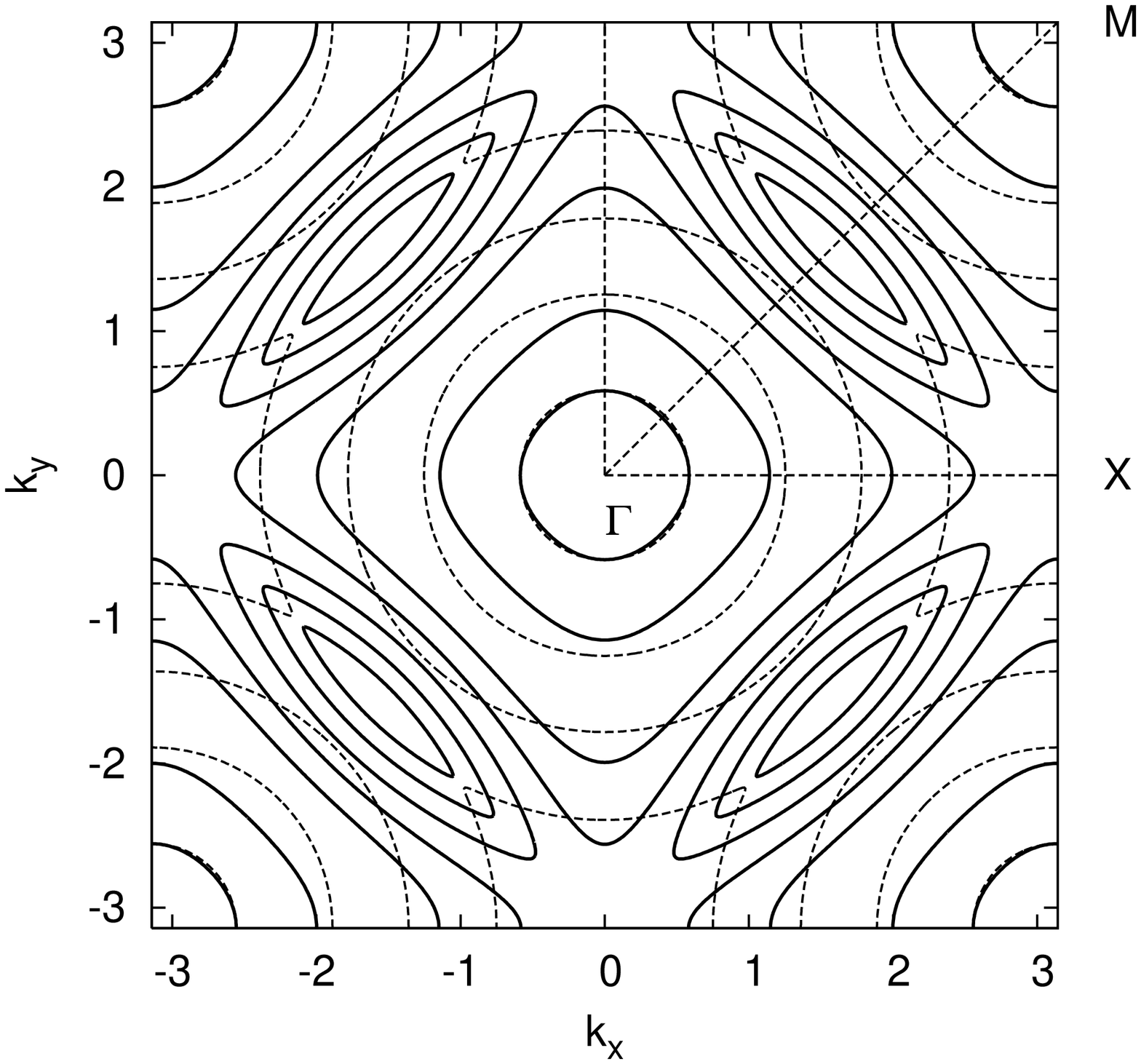}
\end{minipage}\\
\begin{minipage}[c]{0.49\columnwidth}
\begin{flushleft}
(a)
\end{flushleft}
\end{minipage}
\begin{minipage}[c]{0.49\columnwidth}
\begin{flushleft}
(b)
\end{flushleft}
\end{minipage}
\end{center}
\caption{Static polarizability for the mixed dSC+dDW state
   in momentum space, $F_{\dSC+\dDW} (\bq,0)$,
   Eq.~(\protect\ref{eq:FdSCdDW}) [in eV$^{-1}$, panel (a)], for
   $\Delta_0 = 0.1t = 0.03$~eV, $D_\circ = 0.08t=0.024$~eV, $T=0.01
   t\simeq 35$~K \protect\cite{Zhu:01}, and $\mu=-0.2016$~eV,
   corresponding to a 
   particle-like Fermi line closed around the $\Gamma$ point
   (underdoped regime).
Panel (b) shows the contour plots of the eigenvalue spectrum
   $E_{\bk i}$, Eq.~\protect\ref{eq:Eki}.
}
\label{fig:qdSCdDW}
\end{figure}

Fig.~\ref{fig:qdSCdDW} shows our numerical results for the static
   polarizability $F(\bq,0)$ in the mixed dSC+dDW case,
   Eq.~(\ref{eq:FdSCdDW}).
Representative values of the amplitudes of the dSC and dDW order
   parameters have been taken as in Ref.~\cite{Zhu:01}, \emph{viz.}
   $\Delta_\circ = 0.1 t = 0.03$~eV, 
   $D_\circ = 0.08 t = 0.024$~eV, 
   at $T=0.01t\simeq 35$~K,
for a particle-like Fermi line in the underdoped regime ($t=0.3$~eV,
   $t^\prime /t = 0.3$, $\mu = -0.2016$~eV).
Panel (b) of Fig.~\ref{fig:qdSCdDW} shows the contour plot of the
   eigenvalue spectrum $E_{\bk i}$, Eq.~(\ref{eq:Eki}).
The latter is characterized by pronounced minima near the hot spots at
   $\bQ /2$, which evolve into cone-like nodes in the limit of
   pure dSC ($D_\circ = 0$), or in the very special case $t^\prime =
   \mu =0$ [see Eq.~(\ref{eq:cones})].
Accordingly, Fig.~\ref{fig:qdSCdDW}a for the static polarizability
   $F(\bq,0)$ over the 1BZ is characterized by local maxima at the hot
   spots centered around $\bQ/2$, as is the case in the pure dDW case
   (cf. Fig.~\ref{fig:qdDW}a).
Whereas the precise behavior of the static polarizability $F(\bq,0)$
   is of course determined by the actual amount of dSC+dDW mixing at a
   given temperature and doping, we can conclude that a sizeable dDW
   component manifests itself through the appearance of hole pockets
   centered around $\bQ/2$ in the $\bq$-dependence of $F(\bq,0)$, also
   in the presence of a dSC condensate.

We have next evaluated the spatial dependence of $F(\br,0)$ (not
   shown), by Fourier transforming $F(\bq,0)$ to real space.
While $F(\br,0)$ is characterized by Friedel-like oscillations as
   $|\br|$ increases from the impurity site, as expected
   \cite{Zhu:01,Angilella:02f}, these radial, damped oscillations are
   superimposed by an azymuthal modulation, due to
   the $d$-wave symmetry of the normal state, both in the dPG and in
   the dDW cases.
As a consequence, $F(\br,0)$ is characterized by a checkerboard pattern,
   closely related to the symmetry of the underlying square lattice,
   with local 
   maxima on the nearest neighbor and local minima on the next-nearest
   neighbor sites.
Since these features are common to both the dPG and dDW cases, the
   spatial dependence of the charge density oscillations is not
   directly helpful in distinguishing between the dPG and dDW states.
However, real-space and wave-vector dependences of several quantities
   of interest for STM studies can be easily connected by means of
   Fourier transform scanning tunneling microscopy (FT-STM) techniques
   (see, \emph{e.g.}, Ref.~\cite{Capriotti:03}, and refs. therein).
Such a technique has been proved very effective in detecting
   large-amplitude Friedel oscillations of the electron density on the
   Be(0001) \cite{Sprunger:97,Petersen:98} and Be(10$\bar{1}$0)
   surfaces \cite{Briner:98}, and has been recently discussed in
   connection with experimental probes of fluctuating stripes in the
   HTS \cite{Kivelson:03}.
In particular, FT-STM
   experiments \cite{Sprunger:97,Petersen:98,Briner:98} have evidenced
   the role of correlation and reduced dimensionality in establishing
   such `giant' Friedel oscillations in the electron density.

\subsection{Frequency dependence}
\label{ssec:dynamic}

\begin{figure}[t]
\begin{center}
\begin{minipage}[c]{0.9\columnwidth}
\includegraphics[height=\columnwidth,angle=-90]{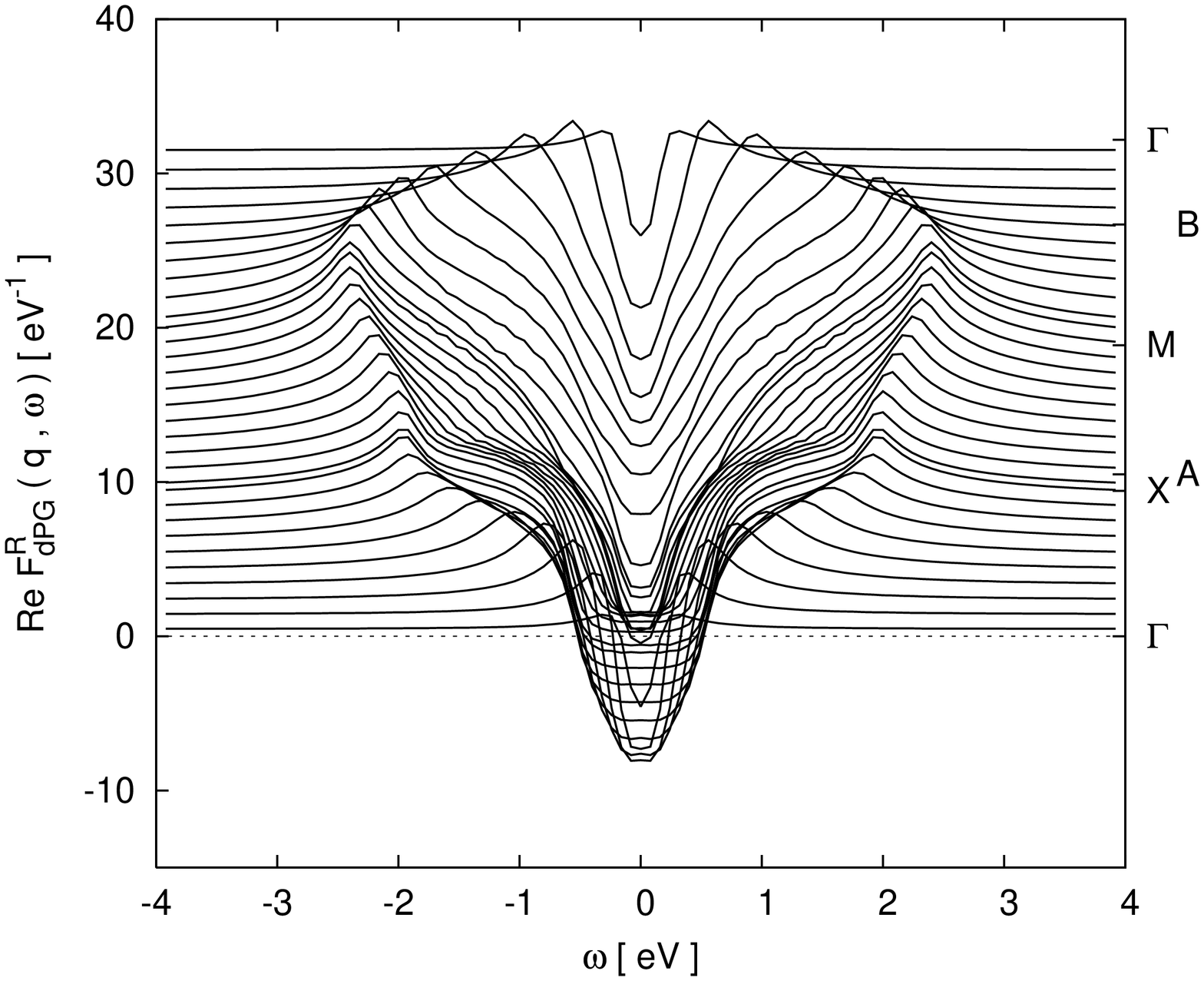}
\end{minipage}
\begin{minipage}[c]{\columnwidth}
\begin{flushleft}
(a)
\end{flushleft}
\end{minipage}
\begin{minipage}[c]{0.9\columnwidth}
\includegraphics[height=\columnwidth,angle=-90]{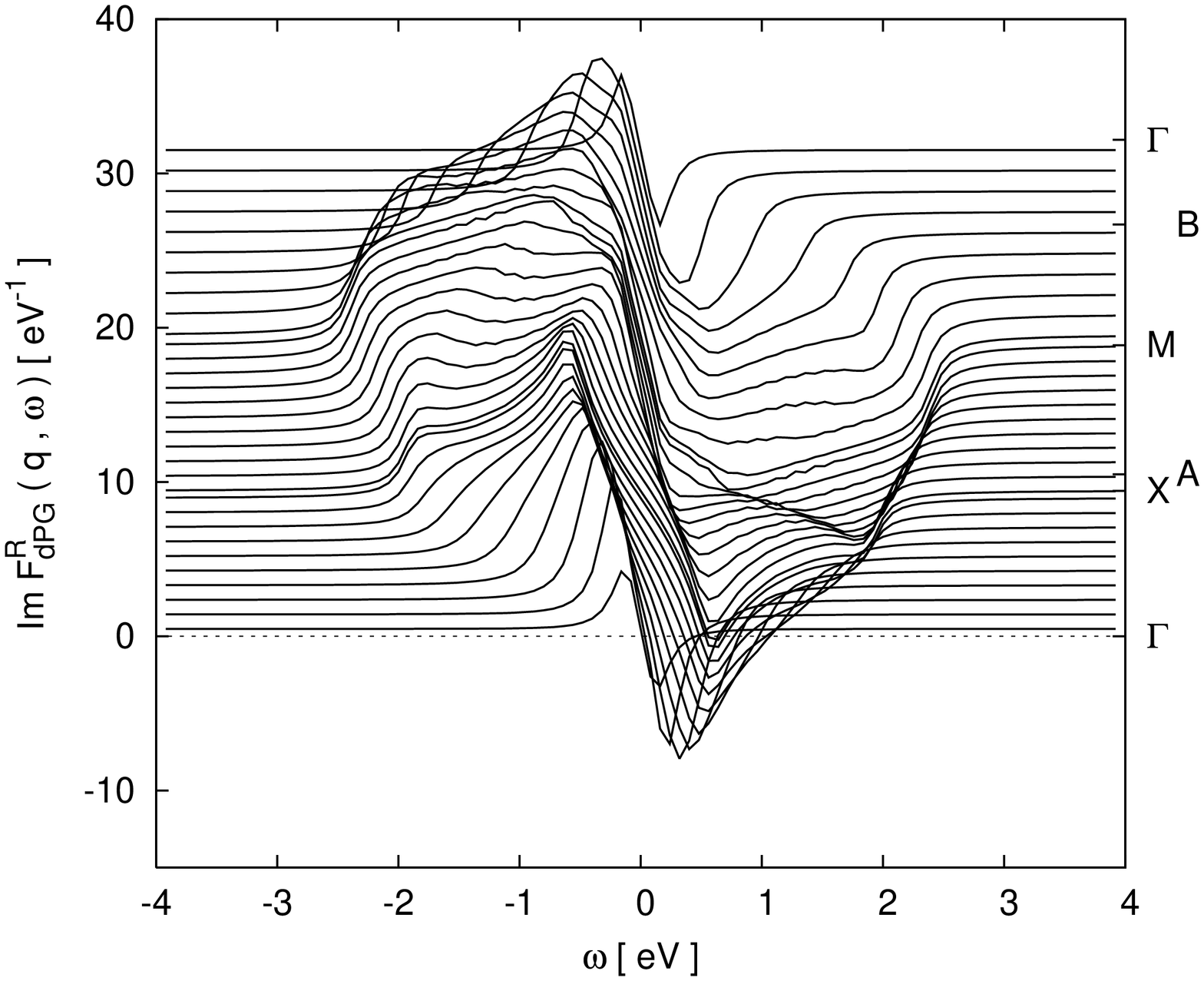}
\end{minipage}
\begin{minipage}[c]{\columnwidth}
\begin{flushleft}
(b)
\end{flushleft}
\end{minipage}
\end{center}
\caption{%
Frequency dependence of the real [panel (a)] and imaginary parts
   [panel (b)] of the retarded
   polarizability, $F^\Ret (\bq,\omega)$, in the
   dPG case, for wave-vector $\bq$ varying along a symmetry contour
   $\Gamma$--$X$--$M$--$\Gamma$ in the 1BZ (see
   Fig.~\protect\ref{fig:qdPG}b).
All curves have been shifted vertically for clarity, by an amount
   proportional to the path length from $\Gamma$ to actual wave-vector
   $\bq$ along such symmetry contour (see right scale).
Dotted line is the zero axis for $F^\Ret (0,\omega)$.
All other parameters as in Fig.~\protect\ref{fig:qdPG}.
}
\label{fig:opg}
\end{figure}

\begin{figure}[t]
\begin{center}
\begin{minipage}[c]{0.9\columnwidth}
\includegraphics[height=\columnwidth,angle=-90]{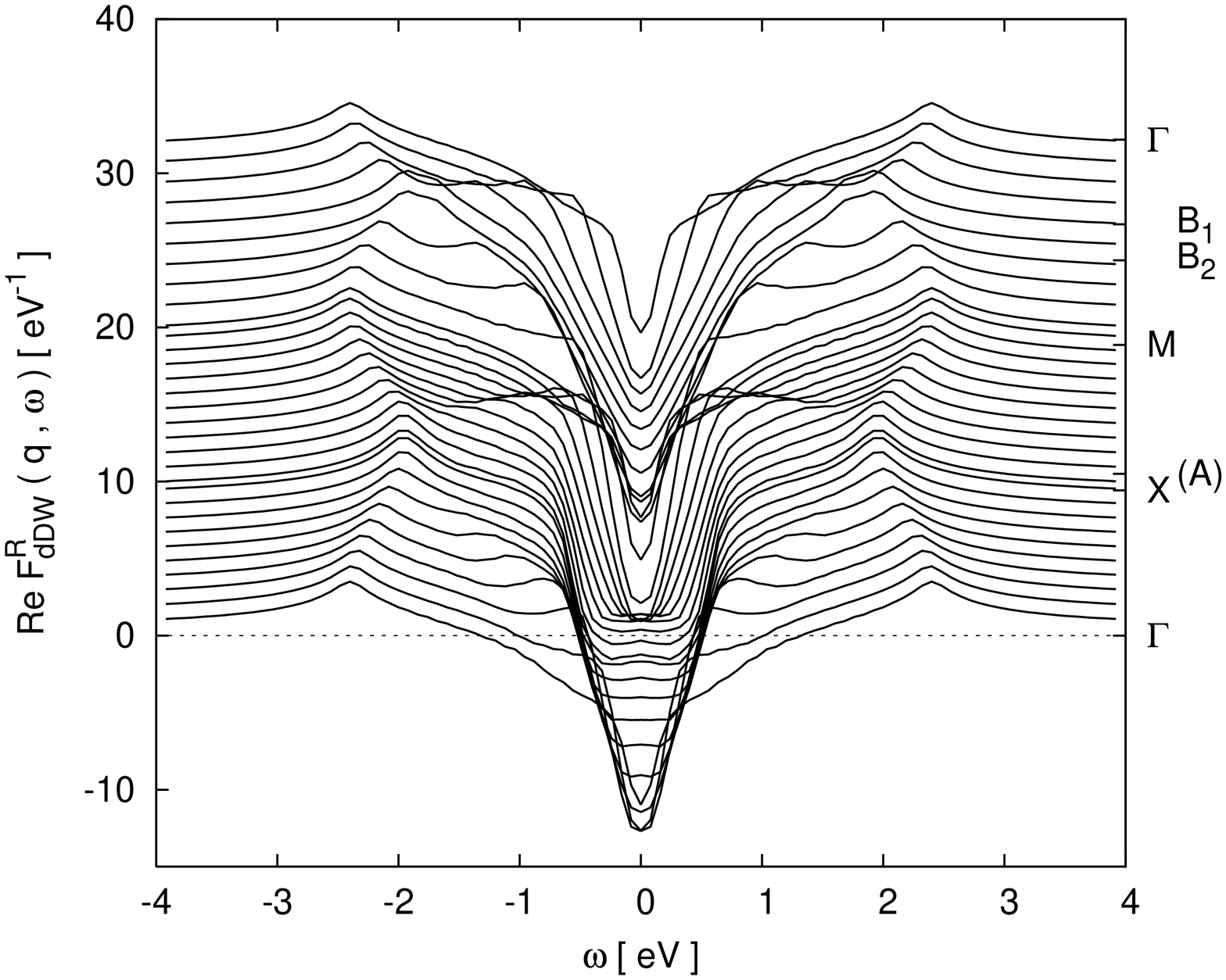}
\end{minipage}
\begin{minipage}[c]{\columnwidth}
\begin{flushleft}
(a)
\end{flushleft}
\end{minipage}
\begin{minipage}[c]{0.9\columnwidth}
\includegraphics[height=\columnwidth,angle=-90]{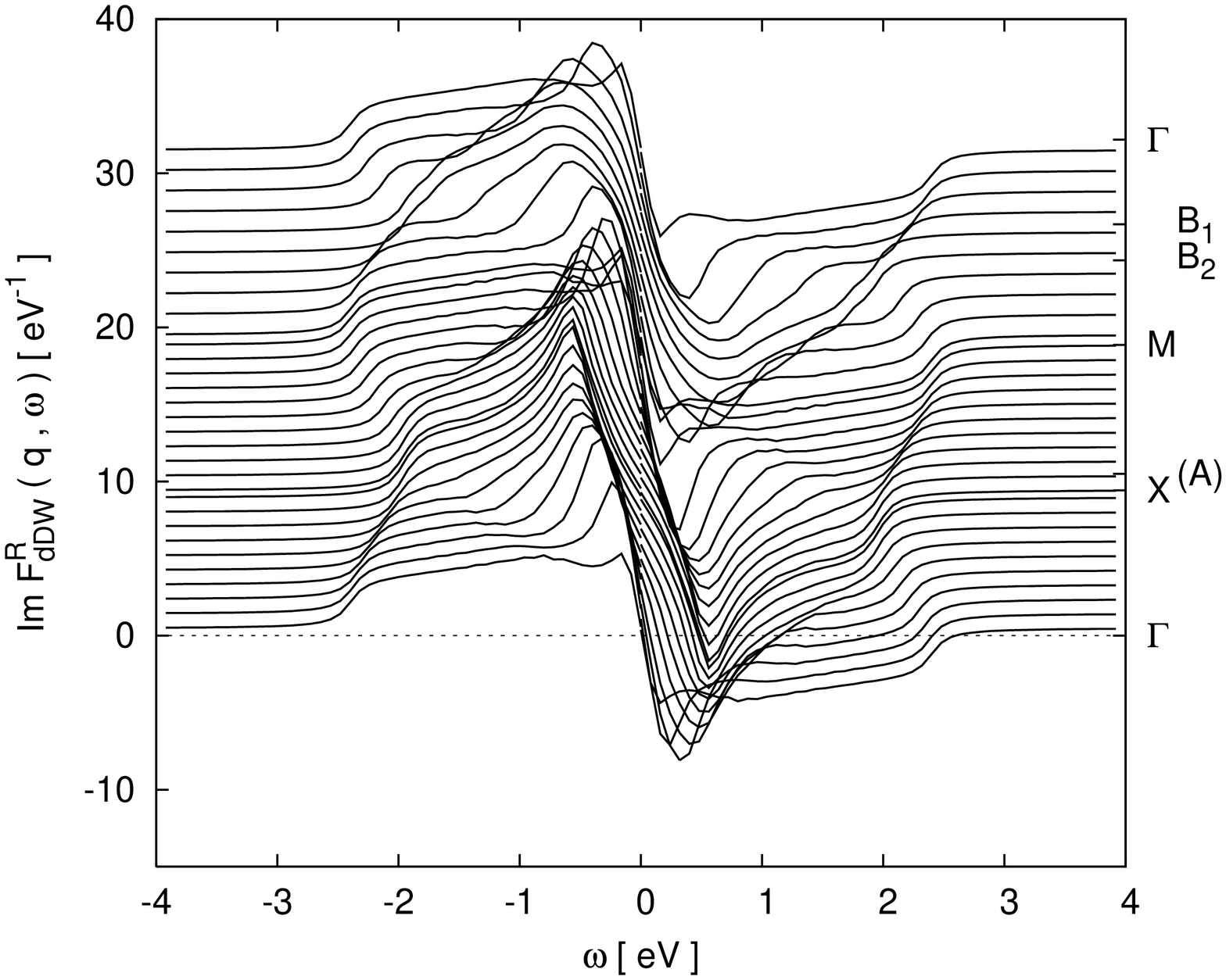}
\end{minipage}
\begin{minipage}[c]{\columnwidth}
\begin{flushleft}
(b)
\end{flushleft}
\end{minipage}
\end{center}
\caption{%
Same as Fig.~\protect\ref{fig:opg}, but for the dDW case.
Special points in the 1BZ are as in Fig.~\protect\ref{fig:qdDW}b.
}
\label{fig:oddw}
\end{figure}

We have next evaluated the frequency dependence of the
   retarded polarizabilities, in both the dPG and the dDW cases.
Figures~\ref{fig:opg} and \ref{fig:oddw} show our numerical results
   for the $\omega$ dependence of $F^\Ret_\dPG
   (\bq,\omega)$ and $F^\Ret_\dDW (\bq,\omega)$, respectively.
Each curve refers to either the real or the imaginary part of $F^\Ret
   (\bq,\omega)$ as a function of 
   $\omega$, for a fixed value of wave-vector $\bq$ along the symmetry
   contour $\Gamma$--$X$--$M$--$\Gamma$ in the 1BZ (see 
   Figs.~\ref{fig:qdPG}b and \ref{fig:qdDW} for its definition).
While $\bq$ runs along such contour, the Fermi line $\xi_\bk =0$ is
   traversed twice (once at point $A$ along $X$--$M$ and once at $B$ along
   $M$--$\Gamma$, for the hole-like Fermi line considered here; see
   Fig.~\ref{fig:qdPG}b), while the
   hole-pocket contour defined by $E_\bk^- = 0$ is traversed twice
   along the $M$--$\Gamma$ line (points $B_1 \equiv B$ and $B_2$ in
   Fig.~\ref{fig:qdDW}b).

As a consequence of the summation over $\bk$ in either the full or
   reduced BZ in Eqs.~(\ref{eq:FdPG}) and (\ref{eq:FdDW}),
   respectively, $\Re F^\Ret (\bq,\omega)$ is an even function of
   $\omega$, while $\Im F^\Ret (\bq,\omega)$ is an odd function of
   $\omega$ in both the dPG and dDW cases.
Therefore, the different contributions of particle and hole states in
   the two cases is averaged out, and no asymmetric peaks in the
   $\omega$ dependence of such quantities are to be expected in the
   dDW case, as is the case for the local density of states
   \cite{Morr:02,MoellerAndersen:03}.

On the other hand, the existence of hole pockets centered around
   $\bQ/2$ in the dDW state is clearly responsible for the
   different $\omega$ dependence of $\Re F^\Ret_\dPG$
   (Fig.~\ref{fig:opg}a) versus $\Re F^\Ret_\dDW$
   (Fig.~\ref{fig:oddw}a), say, as $\bq$ runs along the 
   $\Gamma$--$X$--$M$--$\Gamma$ contours.
While $\Re F^\Ret_\dPG$ is characterized by a single relative maximum
   for $\omega>0$ for all wave-vectors $\bq$ under consideration, $\Re
   F^\Ret_\dDW$ possesses two relative maxima (or a relative maximum
   and a shoulder) for $\omega>0$.
These two maxima tend to merge into a single one for $B_2 \prec \bq
   \prec B_1$, \emph{i.e.} inside the hole pocket, and for $\bq\approx
   A$, \emph{i.e.} at the intersection of the free-particle Fermi line
   with the $X$--$M$ side (Fig.~\ref{fig:oddw}a).
Likewise, the single relative maximum for $\omega>0$ in $\Re
   F^\Ret_\dDW$ shifts towards larger frequencies as $\bq$ runs from
   $\Gamma$ to $X$, is `diffracted' at $A$ along the Fermi line as
   $\bq$ runs from $X$ to $M$, and `bounces back' at $B$, again along
   the Fermi line, as $\bq$ runs from $M$ back to $\Gamma$.
A similar analysis may be performed for $\Im F^\Ret$ in the two cases
   (Figs.~\ref{fig:opg}b and \ref{fig:oddw}b).

As in the static limit, the competition of a sizeable dDW order parameter
   with an underlying dSC condensate does not give rise to
   qualitatively different results in the $\omega$-dependence of the
   polarizability, with respect to the pure dDW case.

One may conclude that, in both the dPG and dDW cases, the evolution
   with $\bq$ of the features in $\omega$ dependence of $F^\Ret
   (\bq,\omega)$ are closely related to the location of wave-vector
   $\bq$ with respect to the Fermi line, and may therefore serve to
   indicate the presence of hole-pockets, as is the case for the dDW
   state.

\section{Conclusions}
\label{sec:conclusions}

Motivated by recent STM experiments around a localized impurity in the
   HTS, we have derived the polarizability (density-density
   correlation function) $F(\bq,\omega)$ for the pseudogap phase, both
   in the precursor pairing scenario and in the $d$-density-wave
   scenario.
Expressions for the same function have been derived also in
   the underdoped regime, characterized by competing dSC+dDW orders.

In the static limit (here defined as the limit of zero external
   frequency for the time-ordered correlation function), the $\bq$
   dependence of $F(\bq,0)$ reflects the $d$-wave symmetry of the
   precursor pairing `pseudogap' or of the dDW order parameter, with an
   azymuthal modulation consistent with a clover-like pattern, as
   expected also for a superconductor with an isotropic band
   \cite{Angilella:02f}.
However, at variance to the dPG case, the $\bq$ dependence of the
   static polarizability in the dDW state clearly exhibits the
   presence of hole pockets, due to the (albeit imperfect) nesting
   properties of the dDW state, with nesting vector $\bQ=(\pi,\pi)$.
Qualitatively similar results to the pure dDW case are obtained also
   in the mixed dSC+dDW, thus showing that hole pockets are a
   distinctive feature of dDW order.
Such a behavior is confirmed by the $\br$ dependence of the static
   polarizability in real space.
A detailed comparison with experimental data for the $\br$-dependence
   of the charge density displacement would of course require a much
   more detailed knowledge of the $\bq$ dependence of the impurity
   potential, here crudely approximated with an $s$-wave Dirac
   $\delta$-function.
In particular, the presence of higher momentum harmonics in the
   impurity potential may break the $d$-wave symmetry of the possible
   correlated or ordered states (dPG, dSC, dDW) here studied.
Also, an extension of the present Born approximation for the impurity
   perturbation, \emph{e.g.} to the $T$-matrix formalism, would afford
   a more reliable comparison with experimental results.

An analysis of the frequency dependence of the retarded polarizability
   $F^\Ret (\bq,\omega)$ reveals that the $\bq$ evolution of the
   features (local maxima or shoulders) in the $\omega$ dependence of
   this function is closely connected with the relative position of
   wave-vector $\bq$ with respect to the Fermi line, and is
   therefore sensitive to the possible presence of hole
   pockets, as is the case for the dDW state. 

\begin{acknowledgments}
We are indebted with P. Castorina, J. O. Fj\ae{}restad, F. E. Leys,
   V. M. Loktev, N. H. March, M. Salluzzo, S. G. Sharapov, F. Siringo,
   D. Zappal\`a for stimulating discussions and correspondence.
\end{acknowledgments}

\appendix

\section{Finite lifetime effects}
\label{app:lifetime}

In order to take into account for finite lifetime effects on the
   linear response function for the pseudogap regime within the
   precursor pairing scenario, we write the diagonal elements of the
   matrix Green's function as
\begin{equation}
\Green_{ii} (\bk, i\omega_n ) = \frac{1}{2\pi} \int d\omega
   \frac{A_{ii} (\bk,\omega )}{i\omega_n - \omega} ,
\end{equation}
where $A_{11} (\bk,\omega) = 2\pi u_\bk^2 \delta(E_\bk -\omega)$,
   $A_{22} (\bk,\omega) = 2\pi v^2_\bk \delta(E_\bk + \omega)$ are the
   appropriate spectral functions for BCS theory.

A finite energy linewidth $\Gamma$ can be attached to the energy state $E_\bk$
   by replacing the $\delta$-functions in the spectral functions
   $A_{ii}$ with broader ones, \emph{e.g.} a Lorentzian function
   $a(\omega) = \frac{1}{\pi} \Gamma/(\omega^2 + \Gamma^2 )$.
Setting
\begin{equation}
A(\bk,\omega) = 2\pi [u_\bk^2 a(E_\bk -\omega) + v_\bk^2 a(E_\bk +
   \omega)] ,
\end{equation}
in the static limit one obtains:
\begin{eqnarray}
F^\Ret_\dPG (\bq,\omega_{\mathrm{ext}}=0) &=&
-\int  \frac{f(\omega)-f(\omega^\prime
   )}{\omega-\omega^\prime} \nonumber\\
&&\times \, \phi(\bq,\omega,\omega^\prime ) \, d\omega \, d\omega^\prime,
\end{eqnarray}
where
\begin{equation}
\phi(\bq,\omega,\omega^\prime ) = \frac{1}{(2\pi)^2} \frac{1}{N}
   \sum_\bk A(\bk,\omega) A(\bk-\bq,\omega^\prime ).
\end{equation}

\section{Polarizability for the \lowercase{d}DW state}
\label{app:poldDW}

In order to derive the analog of the polarizability,
   Eq.~(\ref{eq:Prange}), for the dDW state, we start with considering
   the density-density correlation function:
\begin{equation}
F(\bq,\tau) = -\langle \Tau \rho(\bq,\tau) \rho(-\bq,0) \rangle,
\end{equation}
where $\rho(\bq,\tau)=\sum_{\bk s} c^\dag_{\bq s} (\tau) c_{\bk+\bq s}
   (\tau) $ is the electron density operator, and $T_\tau$ denotes
   ordering with respect to the imaginary time $\tau$.
Application of Wick's theorem then yields
\begin{eqnarray}
F(\bq,\tau) &=& \sum_{\genfrac{}{}{0pt}{2}{\bk\bk^\prime}{s s^\prime}}
\langle \Tau c_{\bk+\bq s} (\tau) c^\dag_{\bk^\prime s^\prime} (0)
   \rangle
\langle \Tau c_{\bk^\prime -\bq s^\prime} (0) c^\dag_{\bk s} (\tau)
   \rangle \nonumber\\
&&-\langle \rho(\bq,0)\rangle \langle \rho(-\bq,0)\rangle ,
\label{eq:app:derdDW:1}
\end{eqnarray}
the last term being a constant with respect to $\tau$, which can be
   neglected in Fourier transforming to the Matsubara frequency
   domain.
In the dDW state, the contributions of terms like Eq.~(\ref{eq:dDWop})
   should be explicitly considered.
Therefore, we make use of the identity
\begin{equation}
\sum_\bk f_\bk = {\sum_\bk}^\prime (f_\bk + f_{\bk + \bQ} ),
\end{equation}
for the summations on both $\bk$ and $\bk^\prime$ in
   Eq.~(\ref{eq:app:derdDW:1}), where the prime 
   restricts the summation to wave-vectors $\bk$ belonging to 
   the reduced (magnetic) Brillouin zone.
After Fourier transforming to the Matsubara frequency domain, one
   eventually has:
\begin{eqnarray}
F_\dDW (\bq,i\omega_\nu ) &=&
\frac{1}{\beta} \sum_{\omega_n} \frac{1}{N} {\sum_\bk}^\prime
[
\Green_{11} (\bk,i\omega_n )
+
\Green_{12} (\bk,i\omega_n )
\nonumber\\
&&\!\!\!\!\!\!\!\!\!\!\!\!\!\!\!\!\!\!\!\!\!\!\!\!\!\!\!\!\!\!\!\!\!\!
+
\Green_{21} (\bk,i\omega_n )
+
\Green_{22} (\bk,i\omega_n )
]\nonumber\\
&&\!\!\!\!\!\!\!\!\!\!\!\!\!\!\!\!\!\!\!\!\!\!\!\!\!\!\!\!\!\!\!\!\!\!\!\!\!\!\!\!\times
   \, [
\Green_{11} (\bk-\bq,i\omega_n -i\omega_\nu )
+
\Green_{12} (\bk-\bq,i\omega_n -i\omega_\nu )
\nonumber\\
&& \!\!\!\!\!\!\!\!\!\!\!\!\!\!\!\!\!\!\!\!\!\!\!\!\!\!\!\!\!\!\!\!\!\!
+
\Green_{21} (\bk-\bq,i\omega_n -i\omega_\nu )
+
\Green_{22} (\bk-\bq,i\omega_n -i\omega_\nu )
],\nonumber\\
\label{eq:app:FdDW}
\end{eqnarray}
where $\Green_{ij}$ are the entries of $\Green_\dDW
   (\bk,i\omega_n )$ in Eq.~(\ref{eq:GdDW}).
The last expression can then be cast into the compact matrix form,
   Eq.~(\ref{eq:PrangedDW}), by introducing the constant auxiliary matrix
   $\kappa=\tau_0 + \tau_1 = \begin{pmatrix} 1 & 1 \\ 1 & 1
   \end{pmatrix}$.
Equation~(\ref{eq:app:FdDW}) simplifies further, by observing that
   $\Green_{12} = -\Green_{21}$ and that $\Green_{11} + \Green_{22} =
   (i\omega_n - E_\bk^+ )^{-1} + (i\omega_n - E_\bk^- )^{-1}$.

\begin{small}
\bibliographystyle{apsrev}
\bibliography{a,b,c,d,e,f,g,h,i,j,k,l,m,n,o,p,q,r,s,t,u,v,w,x,y,z,zzproceedings,Angilella}
\end{small}

\end{document}